\documentclass[aps,prb,twocolumn,superscriptaddress,showpacs]{revtex4}
\usepackage{bm}
\usepackage{amsmath}
\usepackage{amssymb}
\usepackage{graphicx}
\graphicspath{{img_majorana_article_v9/}}  
\usepackage{color}

\begin{document}

\title{Low energy anomalies in electron tunneling through  strongly asymmetric Majorana nanowire}
\author{A.D. Shkop}
\affiliation{B. Verkin Institute for Low Temperature Physics and
Engineering of the National Academy of Sciences of Ukraine, 47
Nauki Ave., Kharkov 61103, Ukraine}

\author{A.V. Parafilo}
\affiliation{The Abdus Salam International Centre for Theoretical Physics,Strada Costiera 11, I-34151 Trieste, Italy }

\author{I.V. Krive}
\affiliation{B. Verkin Institute for Low Temperature Physics and
Engineering of the National Academy of Sciences of Ukraine, 47
Nauki Ave., Kharkov 61103, Ukraine}
\affiliation{Physical Department, V. N. Karazin National University, Kharkov 61077, Ukraine}

\author{R.I. Shekhter}
\affiliation{Department of Physics, University of Gothenburg, SE-412 96 G$\ddot{o}$teborg, Sweden}

\begin{abstract}
Electron transport through Majorana nanowire with strongly asymmetric couplings to normal metal leads is considered. In three terminal geometry (electrically grounded nanowire) it is shown that the presence of unbiased electrode restores zero-bias anomaly even for strong Majorana energy splitting. For effectively two-terminal geometry we show that electrical current through asymmetric Majorana junction is qualitatively different from the analogous current through a resonant (Breit-Wigner) level.
\end{abstract}


\maketitle
\section{Introduction}
Last years Majorana fermions attract a great attention in solid state physics. Firstly predicted by E. Majorana as a fermion particle that coincides with its own antiparticle, Majorana fermions reappeared in condensed matter in the form of Majorana bound states (MBS)-spinless zero-energy subgap edge states in topological superconductors (see e.g.[\cite{Fu}] or review [\cite{Alicea}]), useful for fault-tolerant quantum computation [\cite{Kitaev}]. By definition creation and annihilation operators of MBS coincide, $\gamma_j^{\dag}=\gamma_j$. Being a "half" of a Dirac fermion (its hermitian and anti-hermitian parts), Majorana fermions obey a Clifford algebra, $\lbrace\gamma_i,\gamma_k \rbrace=2\delta_{ki}$. Two MBS localized on the opposite sides of topological superconductor form a highly nonlocal Dirac fermion, $c=(\gamma_1+i\gamma_2)/2, \;c^2=(c^{\dag})^2=0$. This nonlocality leads to unusual electron transport through Majorana bound states. In particular, electron tunneling in Majorana systems could be very different from resonant level electron tunneling even in the case when Majorana hybridization $\varepsilon_M\gamma_1\gamma_2$ ($\varepsilon_M$ is Majorana splitting energy) is taken into account and MBS are splitted into two fermion levels. The presence of substrate superconductor introduces additional (Andreev) channel of electron tunneling and supports electron hole symmetry. Both those properties result in electron tunneling through MBS which strongly differs from ordinary resonant electron tunneling described by Breit-Wigner transmission probability.

Many efforts were spent to theoretically treat these topological modes and distinguish them from "ordinary" excitations in experiment which could mimic the properties of MBS (see e.g.review [\cite{Alicea}]). A promising venue in experimental observation of Majorana fermion is the tunneling experiments where electrons tunnel through MBS which provides the only possible channel for a subgap electrical current at low bias voltages. 

It is already known that Majorana fermions lead to a new transport phenomena - resonant Andreev reflection which manifested in zero bias peak in differential conductance for normal metal/topological superconductor junction [\cite{Lee}]. Although various properties of electron tunnel transport through Majorana bound states have been already studied for two-terminal [\cite{Leijnse}\;\cite{flensb}] and three-terminal [\cite{Zazunov}\;\cite{Lopez}\;\cite{You}] devices, we can add to this knowledge new results concerning specific properties of asymmetric Majorana tunnel junction with strongly different coupling strengths to the normal metal leads.

For this reasons we consider experimental setup (see Fig.1) where an electrically grounded nanowire (i.e. 1D wire on top of s-wave superconductor) is tunnelly coupled to a fixed normal metal electrode (L-electrode) and to a movable tip of scanning microscope (R-electrode). In real experiment Majorana bound states are supposed to be hosted at the ends of semiconducting wire on a top of ordinary s-wave superconductor when proximity effect, strong spin-orbit interaction and external magnetic field work together to form effectively spinless regime of electron transport deep inside the superconducting gap.

Our purpose here is to study transmission properties of topological superconductor with two Majorana modes weakly coupled to the normal metal leads. For electrically grounded superconductor the currents through left (L) and right (R) tunnel contacts in the general case of asymmetric junction ($\Gamma_L\ne\Gamma_R,\Gamma_{L,R}$ are the coupling energies) are different even for equal biases $\mu_L=\mu_R<<\Delta$ ($\Delta$ is the superconducting gap). Each current depends both on $\Gamma_L$ and $\Gamma_R$ if Majorana splitting energy $\varepsilon_M\ne0$. For this junction the linear conductances $G_{\alpha}$ , ($\alpha=L, R$) at low temperatures and $\varepsilon _M=0$  reach maximum value $2e^2/h$, exhibiting zero-bias anomaly in the differential conductance (factor $2$ is due to the contribution of Andreev tunneling) just like when $\Gamma_{L(R)}\rightarrow0$ (see Ref. [\cite{flensb}]). For $\varepsilon_M\ne0$ linear conductances are always finite $G_{\alpha}\ne0$ when both coupling energies $\Gamma_L,\;\Gamma_R$ are finite. In the limit $\Gamma_{L(R)}\rightarrow0$, $\varepsilon_M\ne0$ the linear conductance vanishes, $G_{R(L)}\rightarrow0$. We show that for strongly asymmetric junction $\Gamma_L<<\Gamma_R$ and for finite Majorana energy splitting $\varepsilon_M$ in the range $\Gamma_L<<\varepsilon_M<<\Gamma_R$ the presence of the second MBS at the right end of the Majorana nanowire coupled to the unbiased R-electrode restores zero-bias anomaly in the differential conductance of the left contact. 

In the transport regime when Majorana nanowire is electrically isolated it is shown that electron current through a strongly asymmetric Majorana junction qualitatively differs from the analogous current through Breit-Wigner resonant level.

\begin{figure}[h]
\includegraphics[scale=1]{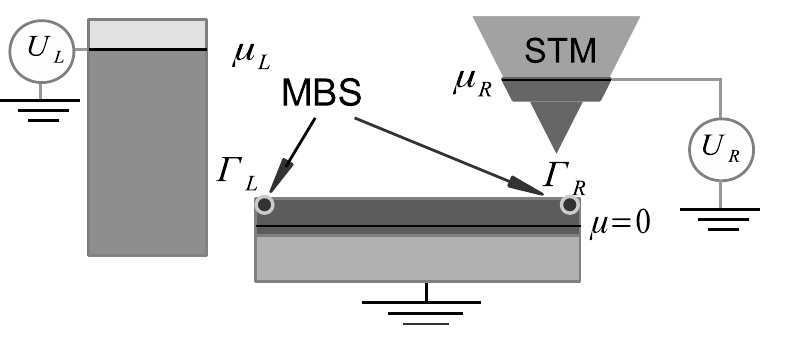}
\caption{A schematic picture of Majorana nanowire with controllable coupling to the leads. Tip of scanning tunneling microscope (STM) at the right end of nanowire enables one to vary the coupling strenght $\Gamma_R$. Electrical potentials of the leads $\mu_{\alpha}=eV_{\alpha},\;\alpha=L,R$ are counted from the electrical potential $\mu=0$ of the electrically grounded topological superconductor.}
\end{figure}

\section{Equations of motion and partial currents}
At first we calculate electric currents at three terminal system consisted of two metal leads and an electrically grounded Majorana nanowire.

The full Hamiltonian is given by three terms
$H=\sum\limits_{\alpha}H_{\alpha}+H_M+H_t$,
where $H_{\alpha}=\sum\limits_k \varepsilon_{\alpha, k} c^{\dag}_{\alpha
k}c_{\alpha k}$ 
is the Hamiltonian of normal leads with 
$c_{\alpha
k}(c_{\alpha k}^{\dag})$ being the electron annihilation (creation)
operator for the $\alpha$ lead (L or R), quantum wire with Majorana edge states is described by effective low energy Hamiltonian
$
H_M=(i/2)\varepsilon_M\gamma_1\gamma_2,
$
which follows from Kitaev toy model [\cite{Kitaev}], here $\varepsilon_M\propto\exp(-L/\xi_0)$ is the splitting between
two zero-energy states  ($L$ is the length of the Majorana quantum
wire and $\xi_0$ is the superconducting coherence length), and $H_t$ is the tunnel Hamiltonian.

The tunnel Hamiltonian describing coupling between $\alpha=L/R$ lead and topological superconductor is
\begin{eqnarray}\label{tunnel}
 H_t=\sum_{\alpha, k} V_{\alpha k} c_{\alpha k}\gamma_{\alpha}+h.c.,
\end{eqnarray}
here $\gamma_{L(R)}=\gamma_{1(2)}$, $V_{\alpha k}$ is the effective amplitude of tunneling which appears due to projection of superconductor electron-field operator onto the manifold of Majorana states, thus tunnel couplings are characterized by energy level width $\Gamma_\alpha=\sum\limits_k 2\pi \delta(\varepsilon-\varepsilon_{\alpha k})|V_{\alpha k}|^2$ (see [\cite{Alicea}\; \cite{flensb}]).

The current operator in the $\alpha$ lead reads ($\hbar=1$)
\begin{equation} \label{current equation}
I_{\alpha}(t)=-e\sum_{k}\frac{d c^{\dag}_{\alpha k}c_{\alpha
k}}{dt}=2e\sum_{k}Im \left(V_{\alpha k}c_{\alpha
k}\gamma_{\alpha}\right).
\end{equation}

By solving the Heisenberg equation of motion for 
$c_{\alpha k}(t)$ one finds
\begin{eqnarray}\label{solv c}
c_{\alpha k}(t)=c_{\alpha k}e^{-i\varepsilon_{\alpha
k}t}
-iV^{\ast}_{\alpha
k}\int_{-\infty}^{t}e^{-i\varepsilon_{\alpha
k}(t-t')}\gamma_{\alpha}(t')dt'.
\end{eqnarray}

Then after substitution it into Heisenberg equation for Majorana operators
\begin{equation}
•\dot{\gamma}_{\alpha}(t)=i[H,\gamma_{\alpha}]
\end{equation}
we obtain matrix equation for them
\begin{equation}
\left(\begin{array}{c}\dot{ \gamma}_L(t) \\
                       \dot{\gamma}_R(t)\end{array}\right)=\left(\begin{array}{cc} -2\Gamma_L & \varepsilon_M \\

                          -\varepsilon_M&
                          -2\Gamma_R\end{array}\right)\left(\begin{array}{c} \gamma_L(t) \\
                       \gamma_R(t)\end{array}\right)+\left(\begin{array}{c} \xi_L(t) \\
                          \xi_R(t)\end{array}\right),
\end{equation}
where
\[\xi_{\alpha}(t)=-2i\sum\limits_k V_{\alpha k}c_{\alpha k}e^{-i\varepsilon_{\alpha k}t}+h.c.\]
After straightforward calculation one finds Majorana operators
\begin{eqnarray}
\gamma_L(t)=&&\sum_k V_{L k}c_{L k}e^{-i\varepsilon_{L
k}t}\dfrac{\varepsilon_{Lk}+2i\Gamma_R}{\Delta_L}+\nonumber \\ 
&&+i\sum_k V_{R k}c_{R k}e^{-i\varepsilon_{R
k}t}\dfrac{\varepsilon_M}{\Delta_R}+h.c.\label{gamma1}
\end{eqnarray}

\begin{eqnarray}
\gamma_R(t)=&&\sum_k V_{R k}c_{R k}e^{-i\varepsilon_{R
k}t}\dfrac{\varepsilon_{Rk}+2i\Gamma_L}{\Delta_R}+ \nonumber \\ 
&&+i\sum_k V_{L k}c_{L k}e^{-i\varepsilon_{L
k}t}\dfrac{\varepsilon_M}{\Delta_L}+h.c.\label{gamma2}
\end{eqnarray}
Here $2\Delta_{\alpha}=[\varepsilon_{\alpha k}+i(\Gamma_L+\Gamma_R)]^2+(\Gamma_L-\Gamma_R)^2-\varepsilon_M^2$.
Now with the help of Eqs.(\ref{current equation}),(\ref{solv c}),(\ref{gamma1}),(\ref{gamma2}) it is easy to get desired expression for the average currents $I_{\alpha}=<I_{\alpha}(t)>$, where $<...>$ is the thermodynamic average with the Hamiltonian of noninteracting electrons in the leads.
The average current $I_{\alpha}=I(T,\mu_{\alpha})$ reads
\begin{eqnarray}
I_{\alpha}=e\int \limits_{-\infty}^{+\infty}d\omega T_{\alpha}(\omega^2)
\tanh\left(\frac{\mu_{\alpha}-\omega}{2k_BT}\right).\label{current}
\end{eqnarray}
Here $T$ is the temperature, $\mu_{\alpha}=eV_{\alpha}$ is the electric potential counted from the Fermi energy and the transmission coefficient $T_{\alpha}(\omega^2)$ takes the form
\begin{equation}
T_{\alpha}(\omega^2)=\frac{4(4\Gamma_L^2\Gamma_R^2+\Gamma_{\alpha}^2\omega^2+\Gamma_L\Gamma_R\varepsilon_M^2)}{\Delta(\omega^2)},
\end{equation}
where 
\begin{eqnarray}
\Delta(\omega^2)=&&\omega^4+4 \omega^2(\Gamma_L^2+\Gamma_R^2)+(4 \Gamma_L\Gamma_R)^2+\nonumber \\
&&+\varepsilon_M^2(\varepsilon_M^2-2(\omega^2-4\Gamma_L\Gamma_R)).\label{delta}
\end{eqnarray}

\section{Differential conductance. Zero-bias anomaly}

Differential conductance in the low temperature limit for each equally biased lead reads (restoring $\hbar$)
\begin{equation}
G_{\alpha}=\frac{2e^2}{h}T_{\alpha}(\omega=eV)\label{conductance}
\end{equation}
and when $V=0,\;\varepsilon_M=0$ it becomes $2e^2/h=2G_0$.
We see that $I(T,\mu_{\alpha}=0)\equiv0$ for arbitrary tunneling rates $\Gamma_L$ and $\Gamma_R$ as it should be when the leads are not biased with respect to the ground. Notice the appearance for spinless electrons an extra overall factor $2$ in Eq.(\ref{conductance}) and hyperbolic tangent in the current dependence on temperature and chemical potential instead of difference of Fermi distribution functions in the ordinary situation (Landauer-Buttiker formula). Both these features are related to the presence of the substrate superconductor in electron transport through Majorana quantum wire. Factor $2$ is due to appearance of addition channel (Andreev tunneling) in electron transport through MBS. Characteristic temperature and chemical potential dependence in Eq.(\ref{current}) is usual for normal metal-superconductor (MS) junctions. In the limiting case of a single MS contact ($\Gamma_L=0$ or $\Gamma_R=0, \varepsilon_M=0$) our formulae for current and conductance are reduced to the corresponding expression in Ref.[\cite{flensb}].
In general case of asymmetric junction ($\Gamma_L \neq \Gamma_R$) the currents in the left and right contacts are not equal, $I_L\neq I_R$ (see also Ref.[\cite{You}]). It is reasonable to consider the limit when the total current to the ground vanishes, $I_G=I_L+I_R=0$. Then one can speak about definite current from the left to right lead induced by voltage bias $eV$. With the help of our general formulae Eqs.(\ref{current})-(\ref{delta}) we reproduce the expression for the current $I=I_L$ through a symmetric Majorana nanowire derived also in Ref.[\cite{Zazunov}\; \cite{Lopez}]. 
For asymmetric junction and/or asymmetric bias $|\mu_L|\neq|\mu_R|$ the total current to the ground $I_G$ is not zero. Here we consider the dependence of differential conductance  on $\mu_L=eV$ in the case when $\mu_R=0$ ($I_R=0,\; I_G=I_L=I(V)$, see also Ref.[\cite{You}]).It is straightforward to find from our basic equations Eqs.(\ref{current})-(\ref{delta}) the dependence of differential conductance on bias voltage at low temperatures $G(V)=2G_0T_L(\omega=eV)$ .In terms of dimensionless variables $\widetilde V=V/2\sqrt{\Gamma_L\Gamma_R}$ and $\tilde \varepsilon_M=\varepsilon_M/2\sqrt{\Gamma_L\Gamma_R}$  differential conductance $G(V)$ takes the form
\begin{equation}
\dfrac{G(V)}{2G_0}=\dfrac{1+\tilde \varepsilon_M^2+(\Gamma_L/\Gamma_R)e^2\widetilde V^2}{(1+\tilde \varepsilon_M^2)^2+e^2\widetilde V^2[(\Gamma_L^2+\Gamma_R^2)/\Gamma_L\Gamma_R-\tilde \varepsilon_M^2]}\label{normalized}
\end{equation}
Particularly in the linear response $V\rightarrow0$ Eg. (\ref{normalized}) is simplified
\begin{equation}
\dfrac{G}{2G_0}=\dfrac{4\Gamma_L\Gamma_R}{4\Gamma_L\Gamma_R+\varepsilon_M^2}. 
\end{equation}
Thus for $\varepsilon_M<<\sqrt{\Gamma_L\Gamma_R}$ differential conductance is
$G/2G_0\rightarrow1$, while $G(0)=0,$ when $\Gamma_R=0,\;\varepsilon_M>> \Gamma_L$.

It means that zero-bias Majorana signature $G(0)=2G_0$ disappears in a single contact junction if Majorana energy splitting $\varepsilon_M>> \Gamma_L$. Zero-bias peak is re-established for strongly asymmetric double contact junction $\Gamma_R>>\Gamma_L$ and $\varepsilon_M<<\sqrt{\Gamma_L\Gamma_R}$ when the total width of splitted Majorana levels exceeds the level splitting. In general the presence of even unbiased second contact enhances the current at low energies (temperature, bias voltage).

\begin{figure}[h]
\includegraphics[scale=1]{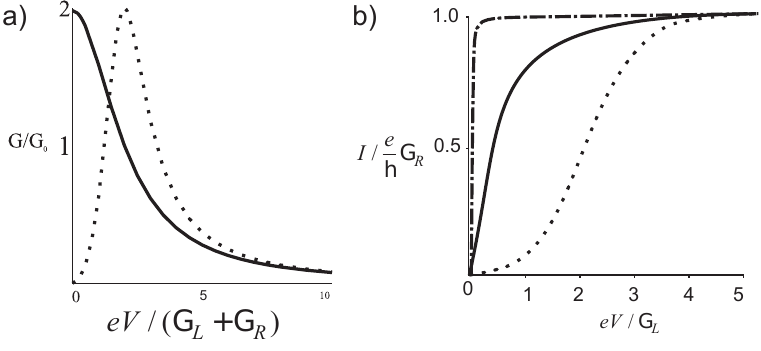}
\caption{a)Differential conductance for electrically grounded Majorana nanowire in units $G_0=e^2/h$ as a function of bias voltage normalized by the total width $\Gamma_L+\Gamma_R$.
(i) solid curve demonstrates the zero-bias anomaly ($\Gamma_R=0$, STM tip is moved to infinity, $\varepsilon_M=0$)
(ii) dotted curve corresponds to the case of strong splitting energy $\varepsilon_M=2\Gamma_L$. Majorana signature $G(V=0)=2G_0$ disappears and conductance peak shifts to nonzero voltages.
When the second contact (right) with high transparency $\Gamma_R>>\Gamma_L$ is introduced one can observe Majorana signature again, line for current dependence in this case coincides with solid line.
b) Current-voltage characteristics of electrically isolated strongly asymmetric Majorana nanowire $\Gamma_R/\Gamma_L=0.001, \varepsilon_M=0\;(dash-dot),\;\varepsilon_M=\Gamma_L\;(solid),\;\varepsilon_M=2\Gamma_L \;(dot)$. In strongly asymmetric system ($\Gamma_L>>\Gamma_R$) without level splitting ($\varepsilon_M=0$), the current saturates at voltages of order of the smallest tunnel width $\Gamma_R$, in contrast to conventional resonant tunneling, thus this dependence is highly nonlinear.} 
\end{figure}

\section{Electrically isolated Majorana nanowire}
Now we consider experimental setup when the superconductor which supports Majorana nanowire is electrically isolated and the current through MBS is induced by the bias voltage $\mu_L-\mu_R=eV$. For a symmetric junction ($\Gamma_L=\Gamma_R=\Gamma$) this problem was studied in Ref.[\cite{Zazunov}\;\cite{Lopez}]. We have seen already that for symmetric electrically grounded junction and for symmetrically biased leads $\mu_L=-\mu_R=eV/2$ (only this case was considered in Ref.[\cite{Lopez}]) the total current to the ground $I_G=I_L+I_R=0$. So the currents through left and right contacts are equal, $|I_L|=|I_R|$. It does not matter whether superconductor is electrically grounded or not.

This strategy can be applied also for asymmetric junction.
Now the equations 
\begin{equation}
\sum_{\alpha=L,R}I_{\alpha}(\mu_{\alpha})=0,\;\;\;\mu_L-\mu_R=eV
\end{equation}
determine electrical potentials $\mu_{\alpha}$ of the leads as a function of bias voltage $V$.
It is evident that for small junction asymmetry $|\Gamma_L-\Gamma_R|<<\Gamma_L+\Gamma_R$ the asymmetry in electrical potentials $\delta V=V_L+V_R$ is small and weakly influences the current. In the opposite limit of strong junction asymmetry (for definiteness we will assume $\Gamma_L>>\Gamma_R$) electrical potentials strongly differ, $|V_L|<<|V_R|$ for all biases $V$ and the current through the electrically isolated Majorana nanowire could be different comparing with the analogous current through resonant (Breit-Wigner) level.

At first we consider low temperature limit $T\rightarrow0$ and a sufficiently long nanowire ($L>>\xi_0$) thus Majorana energy splitting can be neglected. In this case the problem can be easily solved analytically. When $\varepsilon_M=0$ the transmission coefficient $T_{\alpha}$ depends (as it should be) only on its coupling energy strength $\Gamma_{\alpha}$
\begin{equation}
T_{\alpha}=\dfrac{4\Gamma^2_{\alpha}}{\omega^2+4\Gamma^2_{\alpha}}
\end{equation}
and the corresponding currents take a simple form
\begin{equation}
I_{\alpha}(\mu_{\alpha})=\dfrac{2}{\pi}\dfrac{e\Gamma_{\alpha}}{\hbar}\arctan\left(\dfrac{\mu_{\alpha}}{2\Gamma_{\alpha}}\right).\label{current_unbiased}
\end{equation}

For strongly asymmetric junction $\Gamma_L>>\Gamma_R$ the solution of Eq.(\ref{current_unbiased})
is
\begin{equation}
\mu_L\simeq2\Gamma_R\arctan\left(\dfrac{eV}{2\Gamma_R}\right)
\end{equation}
($\mu_R=-eV+\mu_L$) and the current through electrically isolated Majorana nanowire is determined by the corresponding current through the weakest link
\begin{equation}
 •I(V)=\dfrac{2}{\pi}\dfrac{e\Gamma_R}{\hbar}\arctan\left(\dfrac{eV}{2\Gamma_R}\right)\label{current_asymmetric}
 \end{equation} 
According to Eq.(\ref{current_asymmetric}) the current is saturated at $eV>>\Gamma_R$ to the value $I_m=\dfrac{e\Gamma_R}{\hbar}$ which coincides with corresponding maximum current through Breit-Wigner resonant level ($\Gamma_L>>\Gamma_R$). However unlike usual transport where saturation occurs at $eV\simeq\Gamma_{tot}=\Gamma_L+\Gamma_R\simeq\Gamma_L$ (for strongly asymmetric junction) in our case the current reaches its maximum value at a much more lower energies $eV\simeq\Gamma_R$ (see Fig.2,b).

Now we consider the influence of finite Majorana splitting $\varepsilon_M$ on current voltage characteristics. Our calculations show (see Fig.2,b) that "small" values of splitting energy $\varepsilon_M<<\Gamma_L$ weakly influence $I$-$V$ curves evaluated for $\varepsilon_M=0$. When $\varepsilon_M$ is of the order of $\Gamma_L$ the saturation of current curves occurs at energy scale $eV_s\sim\varepsilon_M$ end this $I$-$V$ characteristic resembles the well-known $I(V)$-dependence for electron tunnelling through an asymmetric single-level quantum dot. Specific features of Majorana tunneling disappear.

One can see the characteristic properties of Majorana tunneling also by analysing the temperature dependence of conductance $G(T)$. As it is well known (see e.g. review [\cite{Krive}]) the conductance at resonant tunneling at high temperatures scales as $G\sim \Gamma/T$ (where $\Gamma=\Gamma_L\Gamma_R/(\Gamma_L+\Gamma_R)$) and the crossover temperature from T-independent regime of transport to $1/T$ -scaling is determined by the total level width $\Gamma_t=\Gamma_L+\Gamma_R$. Our calculations show (see Fig. 3) that for strongly asymmetric electrically isolated Majorana nanowire crossover temperature is determined by the weakest coupling and therefore the conductance is strongly suppressed by temperature even at a low temperatures.

\begin{figure}[h]
\includegraphics[scale=1]{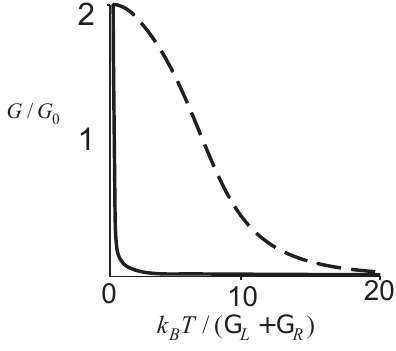}
\caption{Temperature dependence of dimensionless conductance ($G_0=e^2/h$) of electrically isolated Majorana nanowire ($\varepsilon_M=0$).
(i) dashed curve corresponds to symmetric junction $\Gamma_L=\Gamma_R$, (ii) solid curve describes strongly asymmetric junction $\Gamma_R/\Gamma_L=10^{-3}$}.\label{G(T)}
\end{figure}

\section{Conclusions}

In summary we calculated electrical current through Majorana bound states for electrically grounded system end effectively electrically isolated Majorana nanowire. Our aim was to find specific features of electron tunneling in this system in the presence of finite Majorana energy splitting $\varepsilon_M$ wich suppresses zero-bias anomaly in differential conductance. We show that the fingerprints of Majorana states can be easily revealed in tunneling experiments with strongly asymmetric Majorana junction.

We suggested experimental setup where the strenght of MBS coupling to the leads can be controlled with the help of scanning tunneling nicroscope (STM). For three-terminal geometry (electrically grounded Majorana nanowire) it was shown that the presence of unbiased extra electrode strongly coupled to the nanowire increases electric current through Majorana bound states at low bias voltages. In particular in the case when Majorana energy splitting is in the range $\Gamma_L<<\varepsilon_M<<\Gamma_R$ zero-bias anomaly in differential conductance which is suppressed for two-terminal device ($\Gamma_R=0$) is restored when $\Gamma_R$ exceeds $\varepsilon_M$.

Unusual tunneling characteristics of Majorana bound states (MBS) can be observed even in the limit of vanishingly small Majorana energy splitting  $\varepsilon_M\rightarrow0$. It  is known (see e.g.[\cite{flensb}]) that in this case transmission coefficient of electron tunneling through MBS takes the form of Breit-Wigner resonant tunneling probability. Therefore the presence in the system resonant levels at Fermi-energy (in particular, Kondo resonance) can mimic the properties of Majorana fermions. We showed that the tunneling current through electrically ungrounded Majorana nanowire (two-terminal device) with strongly different couplings to the leads is qualitatively distinct from the analogous current through resonant  (Breit-Wigner) level. For sufficiently strong asymmetry the current is saturated at low bias voltages and the measured $I-V$ characteristics will look like a step-function. 

The authors thanks S.I. Kulinich for fruitful discussions. A.S. and I.K. acknowledge financial support from the NAS of Ukraine (grant 4/15-H ). A.P. thanks the Abdus Salam ICTP (Trieste, Italy) for financial support and hospitality.


\begin{thebibliography}{99}

\bibitem{Fu}
 L. Fu and C.L. Kane, Phys. Rev. Lett. {\bf 100}, 096407 (2008).

\bibitem{Alicea}
J.Alicea, Rep. Prog. Phys. {\bf 75}, 076501 (2012). 

\bibitem{Kitaev}
A. Yu. Kitaev, Physics- Uspekhi {\bf44}, 131 (2001).

\bibitem{Akhmerov}
A.R. Akhmerov, J. Nilsson, C.W.J. Beenakker,  Phys. Rev. Lett {\bf 101}, 120403 (2008).

\bibitem{Lee}
K. T. Law, P.A. Lee, and T. K. Ng, Phys. Rev. Lett. {\bf 103}, 237001 (2009). 

\bibitem{Leijnse}
M. Leijnse, K. Flensberg, Phys. Rev.{\bf B 84}, 140501(R) (2011). 

\bibitem{flensb}
K. Flensberg, Phys.Rev.{\bf B 82}, 180516 (2010).

\bibitem{Zazunov}
R. Hutzen, A. Zazunov, B. Braunecker, A. Levy Yeyati, and R. Egger, Phys. Rev. Lett. {\bf 109}, 166403 (2012).

\bibitem{Lopez}
R. Lopez, M. Lee, L. Serra, J. Lim, arXiv:1310.6282(2013). 

\bibitem{You}
Jia-Bin You, Xiao-Qiang Shao, Qing-Jun Tong, A. H. Chan, C. H. Oh, V. Vedral,  J. Phys. Condens Matter {\bf 27(22)}:225302 (2015).

\bibitem{Krive}
I.V. Krive, A. Palevski, R.I. Shekhter, and M. Jonson, Fiz. Nizk. Temp. {\bf 36}, 155 (2010) [Low Temperature Physics {\bf 36}, 119 (2010)]

\end{thebibliography}
\end{document}